# Scalable Solar-Blind Imaging Enabled by Single-Crystalline β-Ga$_2$O$_3$ Membranes on Silicon Backplanes

Xiang Xu[1,2,8], Hong Huang[3,8], Qi Huang[6,8], Hao Wang[2,8], Huaze Zhu[2], Junwei Cao[1,2], Zheng Zhu[7], Yaqin Ma[1,2], Yiyang Xu[1,2], Zhongfang Zhang[4], Yitong Chen[4], Ke Chen[7], Tong Jiang[2,4]*, Bowen Zhu[4,5,6]*, Xiaolong Zhao[3]*, Shibing Long[3], Wei Kong[2,4,5,6]*

[1] School of Materials Science and Engineering, Zhejiang University, Hangzhou, Zhejiang 310027, China.

[2] Department of Material Science and Engineering, School of Engineering, Westlake University, Hangzhou 310000, China.

[3] School of Microelectronics, University of Science and Technology of China, Hefei, Anhui 230026, China.

[4] Zhejiang Key Laboratory of 3D Micro/Nano Fabrication and Characterization, School of Engineering, Westlake University, Hangzhou, Zhejiang 310030, China.

[5] Research Center for Industries of the Future, Westlake University, Hangzhou, Zhejiang 310030, Zhejiang 310030, China.

[6] Westlake Institute for Optoelectronics, Fuyang, Hangzhou, Zhejiang 311400, China.

[7] Center for Neutron Science and Technology, Guangdong Provincial Key Laboratory of Magnetoelectric Physics and Devices, State Key Laboratory of Optoelectronic Materials and Technologies, School of Physics, Sun Yat-Sen University, Guangzhou, Guangdong 510275, China.

[8] These authors contributed equally.

*Email: jiangtong@westlake.edu.cn; zhubowen@westlake.edu.cn; xlzhao77@ustc.edu.cn; kongwei@westlake.edu.cn

# Abstract

Ultrawide-bandgap semiconductors are attractive for solar-blind ultraviolet (UV) detection owing to their intrinsically low noise and high spectral selectivity, yet their deployment in large-area, high-density electronic imaging systems remains limited by a fundamental trade-off between material quality, device speed, and compatibility with high-density planar silicon readout circuits. Here, we report a membrane-enabled integration platform based on transferable single-crystalline β-$Ga_2O_3$ that overcomes these constraints at the system level. By exploiting the weak interplanar bonding of β-$Ga_2O_3$ (100) plane, we obtain wafer-scale freestanding single-crystalline membranes that enable vertically integrated photodiodes with sub-microsecond, non-persistent photoresponse and high UV–visible rejection. Crucially, we introduce a stitching-based membrane assembly strategy that decouples array resolution from the size of the source single-crystalline substrate, allowing high-resolution photodetector arrays to be integrated onto silicon thin-film-transistor backplanes. The modular assembled active-matrix UV imaging arrays exhibit uniform solar-blind response without image lag, in stark contrast to arrays based on amorphous or polycrystalline films. Beyond β-$Ga_2O_3$, this membrane-enabled and stitching-based modular integration strategy provides a general route toward high-speed, high-resolution electronic imaging systems using transferable single-crystalline semiconductors.

# Introduction

High-performance electronic imaging systems demand the simultaneous realization of fast temporal response, high spatial resolution and seamless compatibility with planar readout circuitry[1-4]. These requirements become particularly stringent in the ultraviolet (UV) regime, where applications spanning environmental and industrial monitoring, real-time UV imaging and secure optical communication demand low noise, high spectral selectivity and stable operation under dynamic illumination conditions[5-7]. Ultrawide-bandgap semiconductors offer intrinsic advantages for such applications, including suppressed thermal noise, strong UV–visible discrimination and environmental robustness making them attractive material platforms for advanced UV electronic systems[8-10].

Among ultrawide-bandgap semiconductors, β-$Ga_2O_3$ has attracted increasing interest owing to its large bandgap and favorable electronic properties[11-13]. Substantial progress has been achieved in improving the performance of individual β-$Ga_2O_3$ photodetectors through advances in epitaxial growth[14, 15], defect control[16] and contact engineering[17]. However, translating these material- and device-level advances into large-area, high-density electronic imaging systems remains challenging[18]. In practice, existing UV imaging platforms face a persistent trade-off between material quality, device speed, and compatibility with planar silicon-based active-matrix readout circuits, fundamentally limiting system-level performance[19-21].

Current approaches to β-$Ga_2O_3$-based UV detection can be broadly categorized into two classes. Wafer-bound single-crystalline devices exhibit excellent intrinsic characteristics, including high solar-blind selectivity and low dark current, but are typically implemented in lateral geometries on rigid substrates[22-24]. Such configurations constrain carrier transport length, reduce fill factor and prohibit heterogeneous integration with planar thin-film-transistor (TFT) backplanes when extended to array formats[25-27]. Conversely, large-area UV detector arrays based on polycrystalline or amorphous wide-bandgap films are readily compatible with silicon TFT technology through scalable deposition processes[28, 29]. Yet, their high densities of structural and chemical defects give rise to trap-mediated carrier dynamics, persistent photoresponse, and pronounced image lag, leading to degraded temporal resolution, pixel-to-pixel non-uniformity, and reduced imaging fidelity at the system level[13, 30, 31].

Recent advances in freestanding semiconductor membranes enabled by thin-film lift-off have opened new opportunities to decouple crystal growth from device integration while preserving single-crystalline material quality[32-34]. Access to transferable single-crystalline membranes enables device architectures and heterogeneous integration schemes that are inaccessible to wafer-bound crystals, offering a potential pathway to reconcile intrinsic material performance with system-level integration[35, 36]. Nevertheless, existing demonstrations of membrane-based electronics remain largely confined to single devices or wafer-scale transfers. As a result, array size and pixel resolution are still fundamentally constrained by the dimensions of the source crystal and by conventional wafer-level processing flows, leaving scalable assembly of single-crystalline membranes into high-resolution active-matrix arrays largely unexplored[37, 38].

Here we address these limitations by establishing a membrane-enabled integration platform based on transferable single-crystalline β-Ga$_2$O$_3$ that bridges intrinsic material advantages with scalable electronic integration. By exploiting the weak interplanar bonding of β-Ga$_2$O$_3$ (100) plane[39], we obtain wafer-scale freestanding single-crystalline membranes that enable vertically integrated photodiodes with ultra-fast and non-persistent photoresponse. Crucially, we introduce a stitching-based membrane assembly strategy that decouples array resolution and integration density from the size of the source crystal, allowing high-density photodetector arrays to be integrated onto silicon thin-film-transistor backplanes. Using this approach, we demonstrate active-matrix UV imaging arrays that combine high spectral selectivity, uniform response and the absence of image lag, establishing a scalable route toward high-speed, high-resolution electronic imaging systems based on transferable single-crystalline semiconductors.

## Wafer-scale freestanding single-crystalline β-Ga$_2$O$_3$ membranes

To enable scalable integration of ultrawide-bandgap electronics with planar silicon circuitry, it is essential to obtain single-crystalline semiconductor membranes that are both large-area and transferable. Such a membrane form factor decouples material growth from device fabrication, allowing electronic architectures and system layouts that are inaccessible to wafer-bound single crystals. In this work, we realize freestanding single-crystalline β-Ga$_2$O$_3$ membranes with wafer-scale lateral dimensions, providing a materials platform that is directly compatible with subsequent device integration and array assembly.

Figure 1a illustrates the fabrication process for wafer-scale freestanding single-crystalline β-Ga$_2$O$_3$ membranes. Homoepitaxial β-Ga$_2$O$_3$ layers were grown on unintentionally doped β-Ga$_2$O$_3$ (100) substrates (see Methods)[40], yielding uniform, high-crystallinity films across a 2-inch wafer (Fig. 1b). To enable membrane release, a metal stressor stack consisting of a titanium (Ti) adhesion layer and a copper (Cu) stress layer was deposited on the epitaxial surface. By controlling the intrinsic stress of the metal layer, the epitaxial β-Ga$_2$O$_3$ film could be cleanly exfoliated from the substrate using thermal-release tape, producing a continuous freestanding membrane at the wafer scale (Fig. 1c-d).

The successful exfoliation of the β-Ga$_2$O$_3$ membrane is enabled by the relatively weak interplanar bonding of the (100) orientation, combined with epitaxial stress introduced during homoepitaxial growth[41]. This combination guides crack propagation preferentially along the epitaxial interface, resulting in a self-limited exfoliation process that preserves membrane integrity and thickness uniformity. Unlike conventional mechanical exfoliation from bulk crystals, which often yields films with uncontrolled thickness variations, the present approach produces membranes with well-defined thickness determined by the epitaxial layer (Supplementary Fig. 1-2). Following exfoliation, electrode arrays can be directly fabricated either on the exposed surface of freestanding β-Ga₂O₃ membranes or after membrane transfer onto target substrates, naturally forming vertical device architectures (Fig. 1a-4). This configuration inherently enables vertical carrier transport across the membrane thickness, yielding vertical photodiode structures well suited for UV detection.

The resulting freestanding β-Ga₂O₃ membranes combine wafer-scale uniformity, single-crystalline quality and transferability, addressing a key materials requirement for silicon-compatible ultrawide-bandgap electronics. This membrane platform establishes the foundation for subsequent device fabrication and scalable array integration, as demonstrated in the following sections.

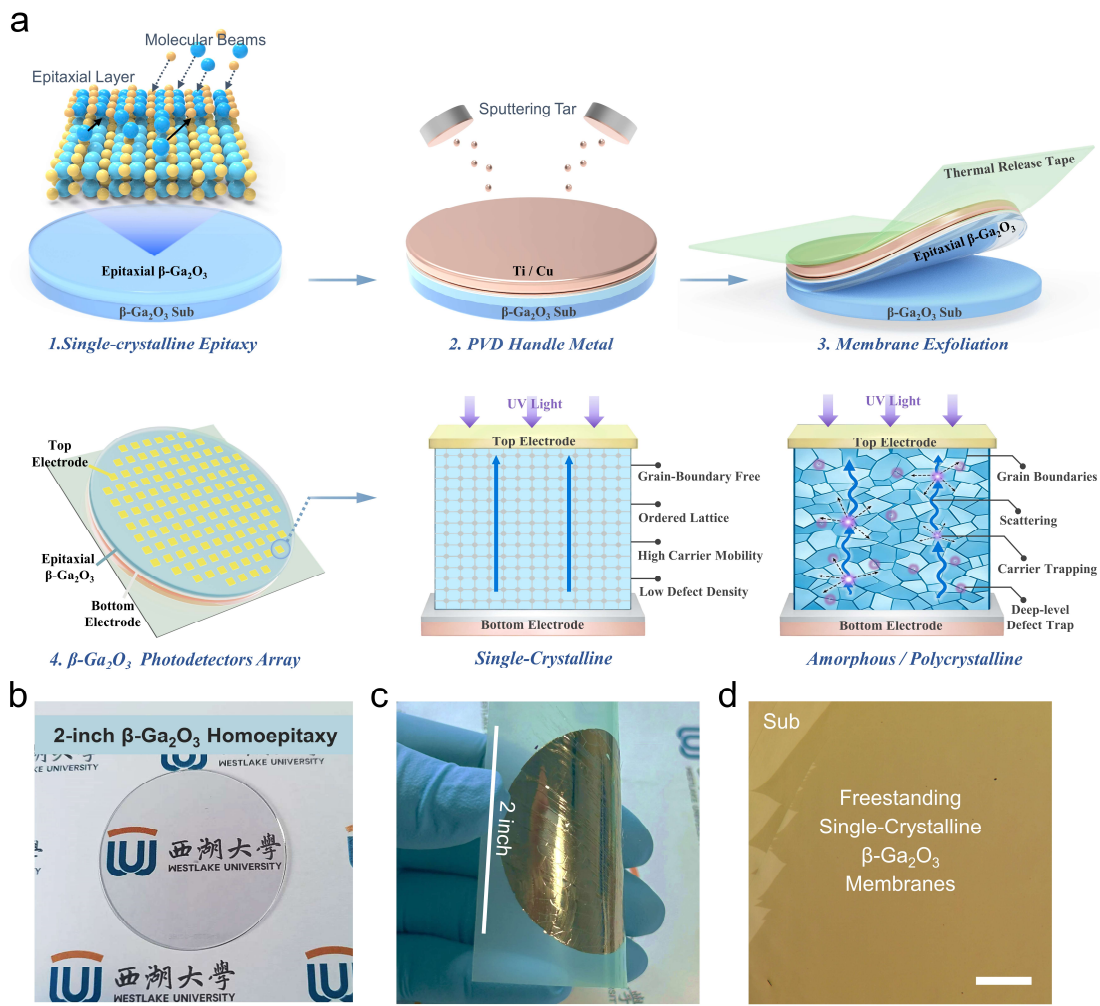

**Fig. 1. Scalable fabrication and transfer of single-crystalline β-Ga₂O₃ membranes. a,** Schematic flowchart for the fabrication of wafer-scale freestanding single-crystalline β-Ga₂O₃ membranes for vertical photodetectors, together with a comparison of carrier transport in single-crystalline and amorphous/polycrystalline β-Ga₂O₃ devices. **b,** Photograph of the 2-inch β-Ga₂O₃ substrate after homoepitaxy growth. **c,** Photograph demonstrating the peeling of the 2-inch single-crystalline β-Ga₂O₃ epitaxial membrane

from the substrate. **d,** Optical micrograph of the single-crystalline β-Ga$_2$O$_3$ membrane after being transferred. Scale bar, 400 μm.

## Structural and electronic properties of single-crystalline membranes.

Material quality critically influences carrier dynamics and photoresponse behavior in wide-bandgap semiconductors[42]. However, single-crystalline semiconductors are difficult to implement at the system level, particularly when integration with silicon-based readout circuits is required, owing to constraints imposed by growth temperature, lattice and thermal mismatch, and wafer-size incompatibility[43, 44]. In practice, this has led to the widespread use of polycrystalline or amorphous wide-bandgap films for system-level integration, despite their compromised material quality. In particular, polycrystalline and amorphous films typically suffer from high densities of structural and chemical defects, leading to trap-mediated carrier dynamics, persistent photoresponse and pixel-to-pixel non-uniformity (see Fig. 1a). These effects fundamentally limit their applicability in high-resolution electronic imaging systems. To evaluate the suitability of the freestanding single-crystalline β-Ga$_2$O$_3$ membranes for such system-level integration, we systematically characterized their structural, chemical, electrical and optical properties and benchmarked them against amorphous Ga$_2$O$_3$ films.

The low crystallographic symmetry of monoclinic β-Ga$_2$O$_3$ makes it intrinsically difficult to suppress in-plane rotational twins when grown on conventional heteroepitaxial substrates with high in-plane symmetry, such as *c*-plane sapphire and 4H-SiC (Supplementary Fig. 3)[45, 46]. By using homoepitaxial growth on β-Ga$_2$O$_3$ (100) substrates, which effectively eliminates twin formation[40], we have successfully fabricated freestanding twin-free single-crystalline β-Ga$_2$O$_3$ membranes. Reflection high-energy electron diffraction (RHEED) patterns recorded along the [010] and [001] azimuths exhibit sharp and streaky features (Fig. 2a), indicating a well-ordered single-crystalline surface with minimal reconstruction or disorder. Electron backscatter diffraction (EBSD) measurements further confirm the twin-free single-crystalline nature of the β-Ga$_2$O$_3$ membranes, which exhibit a single crystallographic orientation, with no randomly distributed colors corresponding to misoriented grains observed along either the [100] or [010] directions (Fig. 2b). Cross-sectional high-angle annular dark-field scanning transmission electron microscopy (HAADF–STEM) imaging reveals an indistinguishable interface between the homoepitaxial layer and the β-Ga$_2$O$_3$ (100) substrate (Fig. 2c), confirming the high crystalline quality of the epitaxial membrane prior to exfoliation. Importantly, the membrane remains structurally intact after lift-off, with no observable interfacial residues or damage, demonstrating that the exfoliation and transfer processes preserve single-crystalline order (Fig. 2d). The high crystalline coherence of the exfoliated membrane is further confirmed by X-ray diffraction characterization (XRD) (see Supplementary Fig. 4 and Note 1).

Chemical purity was examined by X-ray photoelectron spectroscopy (XPS). The O 1s spectrum of the single-crystalline membrane exhibits a sharp and symmetric peak, in contrast to the broadened profile observed for amorphous $Ga_2O_3$ films (Fig. 2e). This indicates a well-defined oxygen bonding with reduced concentrations of suboxide species and defect-related states, consistent with high stoichiometry in epitaxial thin films. Optical absorption spectra (Fig. 2f) reveal that single-crystalline β-$Ga_2O_3$ exhibits a markedly sharper absorption edge and a pronounced blue shift of the optical bandgap compared with its amorphous counterpart. Tauc analysis (inset of Fig. 2f) yields optical bandgaps of approximately 4.89 eV for the single-crystalline membrane and 4.75 eV for the amorphous film. The enlarged bandgap and steeper absorption onset of the single-crystalline β-$Ga_2O_3$ indicate substantially reduced structural disorder and a suppressed density of sub-bandgap states, consistent with its superior crystalline quality. In addition, the single-crystalline membrane exhibits a very high absorption coefficient in the deep-UV region ($\alpha > 1.5 \times 10^5$ $cm^{-1}$). As a result, an ultrathin membrane with a thickness of only 200 nm achieves an absorption of 98.17% at 254 nm (Supplementary Fig. 5).

Ultrafast carrier dynamics were further probed using femtosecond pump–probe spectroscopy (Fig. 2g). The transient reflectivity dynamics of the single-crystalline membrane exhibit rapid relaxation time constants ($\tau_s$=328 ps) with suppressed long-lived components compared with amorphous $Ga_2O_3$ ($\tau_a$=1690 ps), indicating minimized carrier trapping and de-trapping processes. The absence of slow relaxation pathways is particularly important for electronic imaging applications, where persistent photoresponse leads to image lag and degradation of temporal resolution. Moreover, pronounced oscillatory features are clearly resolved exclusively in the single-crystalline β-$Ga_2O_3$ sample and are attributed to coherent phonon oscillations, providing further evidence of its well-defined lattice periodicity and long-range structural order. Consistent with the ultrafast dynamics, single-crystalline β-$Ga_2O_3$ devices exhibit negligible current–voltage hysteresis, whereas amorphous $Ga_2O_3$ shows pronounced hysteresis (Fig. 2h), reflecting suppressed trap-mediated charge dynamics.

Collectively, these characterizations establish that the freestanding single-crystalline β-$Ga_2O_3$ membranes constitute an advantageous electronic materials platform compared with polycrystalline or amorphous films. The combination of high crystallinity, chemical uniformity and suppressed trap-mediated carrier dynamics is essential for stable, fast and uniform operation in large-area electronic arrays.

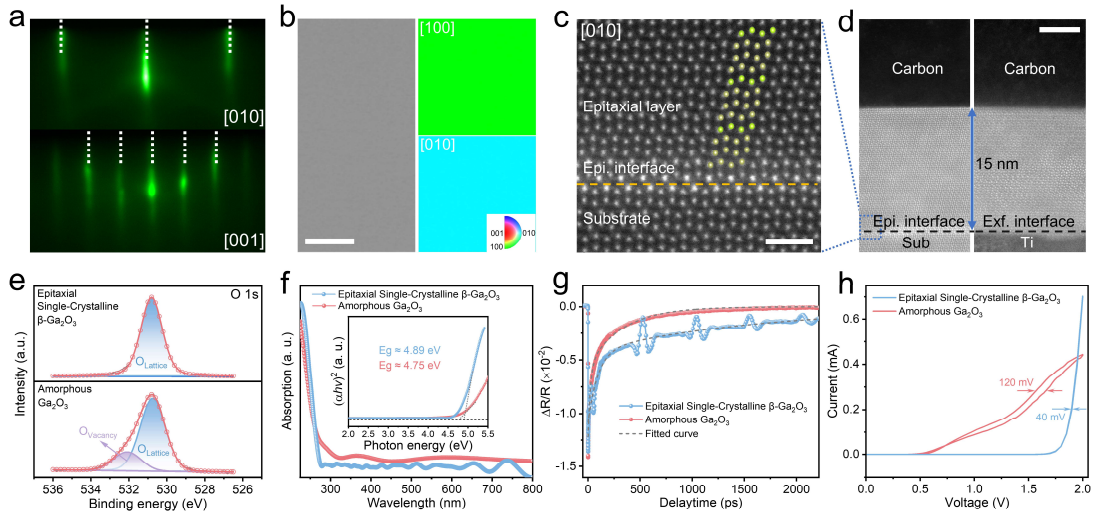

**Fig. 2. Structural integrity and electronic properties of single-crystalline β-Ga$_2$O$_3$ membranes. a,** RHEED patterns of the β-Ga$_2$O$_3$ membranes recorded along the [010] and [001] azimuths, respectively. **b,** SEM image acquired during EBSD measurement (left) and corresponding EBSD orientation maps of the β-Ga$_2$O$_3$ membrane along the (100) and (010) planes (right), confirming uniform single-crystalline orientation. **c,** Cross-sectional HAADF-STEM image of the homoepitaxial β-Ga$_2$O$_3$ membranes grown on a β-Ga$_2$O$_3$(100) substrate. Scale bar: 1 nm. **d,** Cross-sectional STEM images showing a 15-nm β-Ga$_2$O$_3$ epitaxial membrane on a β-Ga$_2$O$_3$(100) substrate (left) and after being exfoliated from the substrate (right). Scale bar: 5 nm. **e,** O 1s XPS spectra of the single-crystalline β-Ga$_2$O$_3$ epitaxial membrane and amorphous Ga$_2$O$_3$ film. **f,** Absorption spectra of the single-crystalline β-Ga$_2$O$_3$ epitaxial membrane and amorphous Ga$_2$O$_3$ film. The inset shows the corresponding Tauc plots of (αhν)$^2$ versus photon energy for direct bandgap analysis. **g,** Transient reflectivity (ΔR/R) dynamics of amorphous Ga$_2$O$_3$ and epitaxial single-crystalline β-Ga$_2$O$_3$ measured by femtosecond pump–probe spectroscopy. **h,** Current–voltage hysteresis curves of single-crystalline β-Ga$_2$O$_3$ and amorphous Ga$_2$O$_3$ photodiode devices measured under dark conditions.

## Membrane-enabled single-crystalline β-Ga$_2$O$_3$ vertical photodetectors.

To evaluate the intrinsic device performance enabled by single-crystalline β-Ga$_2$O$_3$ membranes, we first investigate vertical photodiodes fabricated on individual freestanding membranes. The vertical device geometry minimizes lateral transport and parasitic effects, allowing carrier dynamics to be governed primarily by the membrane thickness and material quality. This configuration provides a stable and reliable platform for single-device benchmarking of temporal response, spectral selectivity and linearity. Importantly, such a vertical architecture is inherently compatible with planar silicon-based processing, providing a device scheme that can be readily extended to wafer-scale integration in subsequent array demonstrations.

Figure 3a illustrates the membrane-enabled vertical β-Ga$_2$O$_3$ photodiode structure employed in this study. A semi-transparent Pt top electrode is deposited on the freestanding β-Ga$_2$O$_3$ membrane to act as a rectifying contact while allowing efficient UV transmission (Fig. 3b, Supplementary Fig. 6), whereas the bottom electrode (Ti/Cu) is formed on the opposite side of the membrane (see Methods). Under UV illumination, photogenerated carriers are separated and collected across the membrane thickness by the built-in electric field and applied bias, enabling efficient vertical transport without reliance on lateral conduction pathways (Fig. 3c).

The resulting vertical photodiodes exhibit clear rectifying behavior with a well-defined turn-on voltage (Fig. 3d), consistent with suppressed trap-assisted transport in the single-crystalline membrane. In contrast, devices based on polycrystalline or amorphous Ga$_2$O$_3$ films typically show broadened turn-on behavior (Supplementary Fig. 7). The vertical single-crystalline membrane-based photodiodes further display well-behaved photodiode characteristics, including low dark current under reverse bias and a light on–off ratio exceeding $5.1 \times 10^5$ at a reverse bias of 2 V, which is more than two orders of magnitude higher than that measured in amorphous Ga$_2$O$_3$ devices under comparable operating conditions (also see Supplementary Fig. 7 and Note 2). Consistent with the suppressed dark current discussed above, Fig. 3e demonstrates a stable photocurrent response over a broad UV intensity range (see Fig. 3f and Supplementary Note 3). The spectral response of the vertical β-Ga$_2$O$_3$ photodiodes is shown in Fig. 3g. The devices exhibit a well-defined cut-off wavelength at 254.6 nm, corresponding closely to the long-wavelength absorption edge of β-Ga$_2$O$_3$. A high ultraviolet-to-visible rejection ratio ($R_{peak}/R_{400\ nm}$) of approximately $10^6$ is obtained, demonstrating excellent solar-blind selectivity. This value is more than three orders of magnitude higher than that measured in sputtered amorphous Ga$_2$O$_3$ devices under comparable conditions (Supplementary Fig. 8 and Note 4). The absence of appreciable response to visible illumination reflects the low density of defect- and disorder-induced sub-bandgap states in the single-crystalline membrane, which effectively suppresses parasitic absorption pathways. Such high spectral selectivity enables filter-free UV photodetection and is particularly advantageous for reducing background noise in practical imaging and sensing applications.

The temporal response of the vertical photodiodes was evaluated under pulsed UV illumination. The rise and decay times of the photocurrent are measured to be 750 ns and 940 ns, respectively, corresponding to a sub-microsecond temporal response (Fig. 3h). This fast response arises from the intrinsically short carrier transit distance enabled by the vertical device geometry. In addition, the high crystalline quality of the β-Ga$_2$O$_3$ membrane suppresses trapping and de-trapping processes as well as bias-induced barrier fluctuations, thereby preventing the formation of long-lived response tails. As a result, the devices exhibit ultra-fast and non-persistent photoresponse, which is critical for high-speed UV detection.

The benchmark plot in Fig. 3i highlights the superior performance of the present devices. By leveraging the low defect density of single-crystalline β-Ga$_2$O$_3$ and the efficient carrier collection inherent to vertical structures, our devices circumvent the typical trade-off between sensitivity and speed. The resulting sub-microsecond response and high UV-to-visible rejection ratio position these membrane-based vertical photodiodes at the forefront of the $Ga_2O_3$ solar-blind detection landscape, as detailed in Supplementary Table 1.

Finally, the operational stability and reliability of the vertical photodiodes were evaluated by periodically switching the UV illumination on and off. As shown in Fig. 3j, the devices maintain stable and reproducible photoresponse characteristics over continuous operation exceeding 4,000 s, indicating robust device performance under repeated cycling. Collectively, these results highlight the advantage of combining high single-crystalline material quality with a vertical device geometry for achieving high-performance and robust photodiode operation.

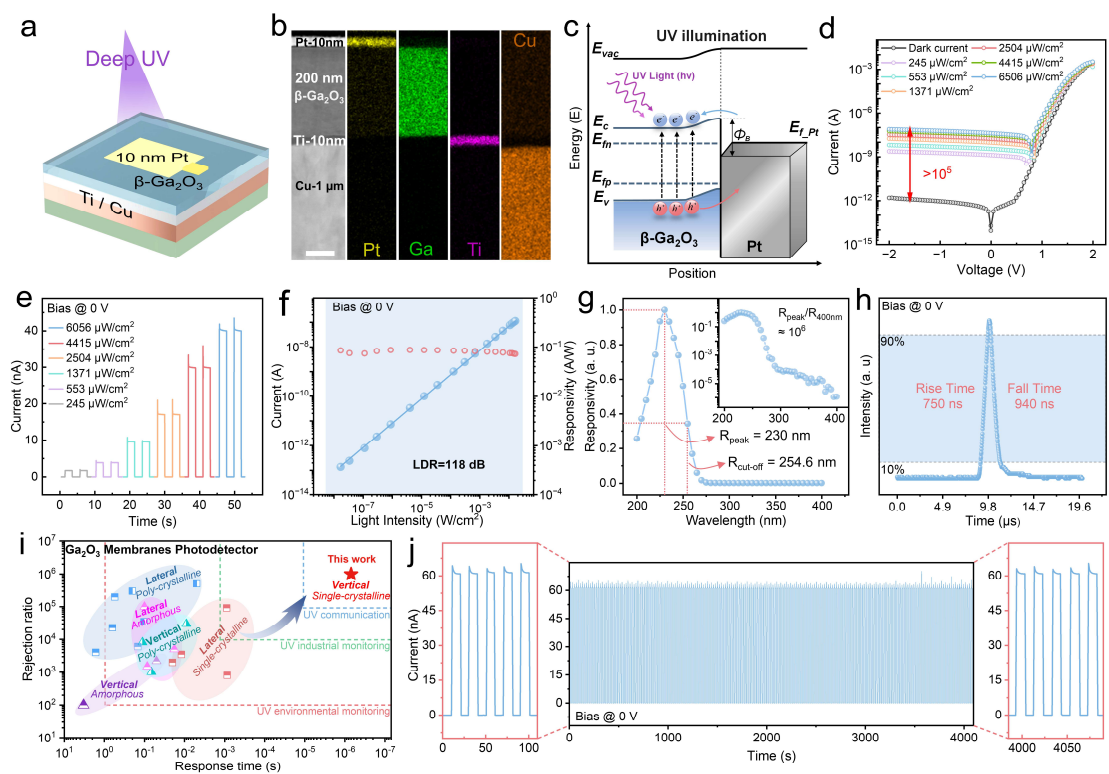

**Fig. 3. Device performance of membrane-enabled vertical β-$Ga_2O_3$ photodiode. a,** Schematic diagram of the photodetectors structure. **b,** Cross-sectional STEM and EDS images of single-crystalline β-$Ga_2O_3$ vertical photodiode. Scale bar, 50 nm. **c,** The energy band schematic diagram of the single-crystalline β-$Ga_2O_3$ photodiode. **d,** I–V characteristic curves of the single-crystalline β-$Ga_2O_3$ photodetectors in the dark and under 254 nm UV illumination. **e,** Time-dependent photodiode of the single-crystalline β-$Ga_2O_3$ photodiode under different light intensities. **f,** The photocurrent and responsivity varying with light intensity at a 0 V bias voltage, demonstrating a linear dynamic range (LDR) of 118 dB. **g,** Normalized spectral response characteristics of the single-crystalline β-$Ga_2O_3$ photodiode. Insert: Exponential coordinate diagram of spectral selective responsivity. **h,** Transient response of the single-crystalline β-$Ga_2O_3$ photodiode excited by a 248 nm pulsed laser. **i,** Benchmark of pure $Ga_2O_3$-based photodetectors, comparing UV–visible rejection ratio and response time as a function of crystal type (single-crystalline, polycrystalline and amorphous) and device geometry (lateral versus vertical). The reference data points are listed in Supplementary Table 1. **j,** Long-term stability characteristics of the single-crystalline β-$Ga_2O_3$ photodiode under a periodical light switch over 4000 s.

# High-resolution photodetector arrays enabled by stitching-based membrane assembly.

While the vertical photodiodes described above establish the intrinsic performance limits of individual membrane-based devices, practical imaging and sensing systems require scalable integration of high-resolution pixels with uniform and stable response. In principle, increasing wafer diameter offers a straightforward route to larger device arrays. In practice, however, wafer scaling in ultrawide-bandgap semiconductors faces compounding limitations associated with crystal growth yield, defect uniformity and cost, particularly for materials such as β-$Ga_2O_3$ that rely on bulk single-crystal substrates. As wafer size increases, maintaining thickness homogeneity, low defect density and mechanical integrity across the entire substrate becomes increasingly difficult, directly impacting array yield and pixel-to-pixel uniformity. Moreover, wafer-bound processing intrinsically couples array size to crystal dimensions, restricting layout flexibility and constraining heterogeneous integration with planar thin-film-transistor backplanes. To overcome these limitations, we employ a stitching-based membrane assembly strategy that decouples array scalability from the dimensions of the source crystal. Such decoupling is particularly advantageous for active-matrix electronics, where backplane technology, rather than crystal growth, ultimately dictates pixel resolution and density, addressing system form factors.

Figure 4a illustrates the stitching-enabled assembly process for constructing large-area photodetector arrays. Individual single-crystalline β-$Ga_2O_3$ membrane segments are first attached to a polydimethylsiloxane (PDMS) stamp, followed by removal of the metal stressor layers, and then sequentially aligned and laminated onto an amorphous silicon thin-film transistor (α-Si TFT) backplane (see Methods and Supplementary Fig. 9). This approach allows multiple membrane tiles to be stitched onto a target TFT backplane of arbitrary dimensions, forming a continuous imaging array without the need for single-crystal substrates of matching size (see Supplementary Fig. 10 and Note 5). Owing to the atomically smooth surface of the exfoliated membranes (Supplementary Fig. 11), intimate interfacial contact with planar substrates can be achieved without chemical bonding, relying on van der Waals interactions. The bottom contact is defined through the underlying substrate, while a semi-transparent Pt top electrode is deposited on the top surface as a common electrode, forming a vertical metal-semiconductor-metal device configuration. This stitching-based assembly represents a general integration strategy for transferable single-crystalline membranes, and can in principle be extended to other membrane-form semiconductors.

A photograph of the completed array is shown in Supplementary Fig. 12. Figure 4b presents an optical micrograph of the 256 × 256 TFT backplane, with a magnified view of the pixel region shown in Fig. 4c, revealing the structures of the switching TFTs and the bottom contacts of the photodetectors. Each pixel consists of a β-$Ga_2O_3$ photodiode connected in series with a TFT, as depicted in the equivalent circuit diagram (Fig. 4d) and the structural schematic (Supplementary Fig. 13). This active-matrix architecture enables row-by-row and column-by-column addressing of individual photodetectors for sequential readout. Array imaging performance was recorded using a real-time acquisition system (Fig. 4e and Supplementary Fig. 14-15).

The advantages of the single-crystalline membrane platform become particularly evident in dynamic imaging measurements. Figures 4f–g compare the imaging behavior of arrays fabricated from single-crystalline β-$Ga_2O_3$ membranes and sputtered amorphous $Ga_2O_3$ films under identical operating conditions. In the single-crystalline membrane-based array, the image rapidly returns to a dark state upon removal of UV illumination, directly reflecting the non-persistent photoresponse observed at the single-device level. The array responds exclusively to UV illumination and produces spatially uniform images. In contrast, arrays based on amorphous $Ga_2O_3$ films exhibit responses to both solar and UV illumination, together with pronounced image lag and residual signals persisting for several seconds after the light is turned off. These artefacts arise from poor spectral selectivity and long-lived trap-mediated carrier dynamics in the amorphous material.

The high-resolution imaging capability of the single-crystalline β-$Ga_2O_3$ photodetector array is further demonstrated by clear visualization of a line-pair chart (~5 lp mm$^{-1}$, limited by the pixel size of 100 × 100 μm²) and a fingerprint motif (Fig. 4h-i). These results confirm the ability of the array to resolve fine structural details and capture complex patterns with high fidelity. In addition, systematic imaging measurements reveal that both the contrast and effective sensitivity of the array are readily modulated by the applied common bias and incident UV power (Supplementary Fig. 16), offering an additional degree of freedom to optimize dynamic range for different imaging scenarios. Taken together, the stitched single-crystalline β-$Ga_2O_3$ photodetector array combines high spectral selectivity, uniform and rapid non-persistent photoresponse, and high spatial resolution, providing a promising platform for fast, high-contrast, solar-blind UV imaging applications.

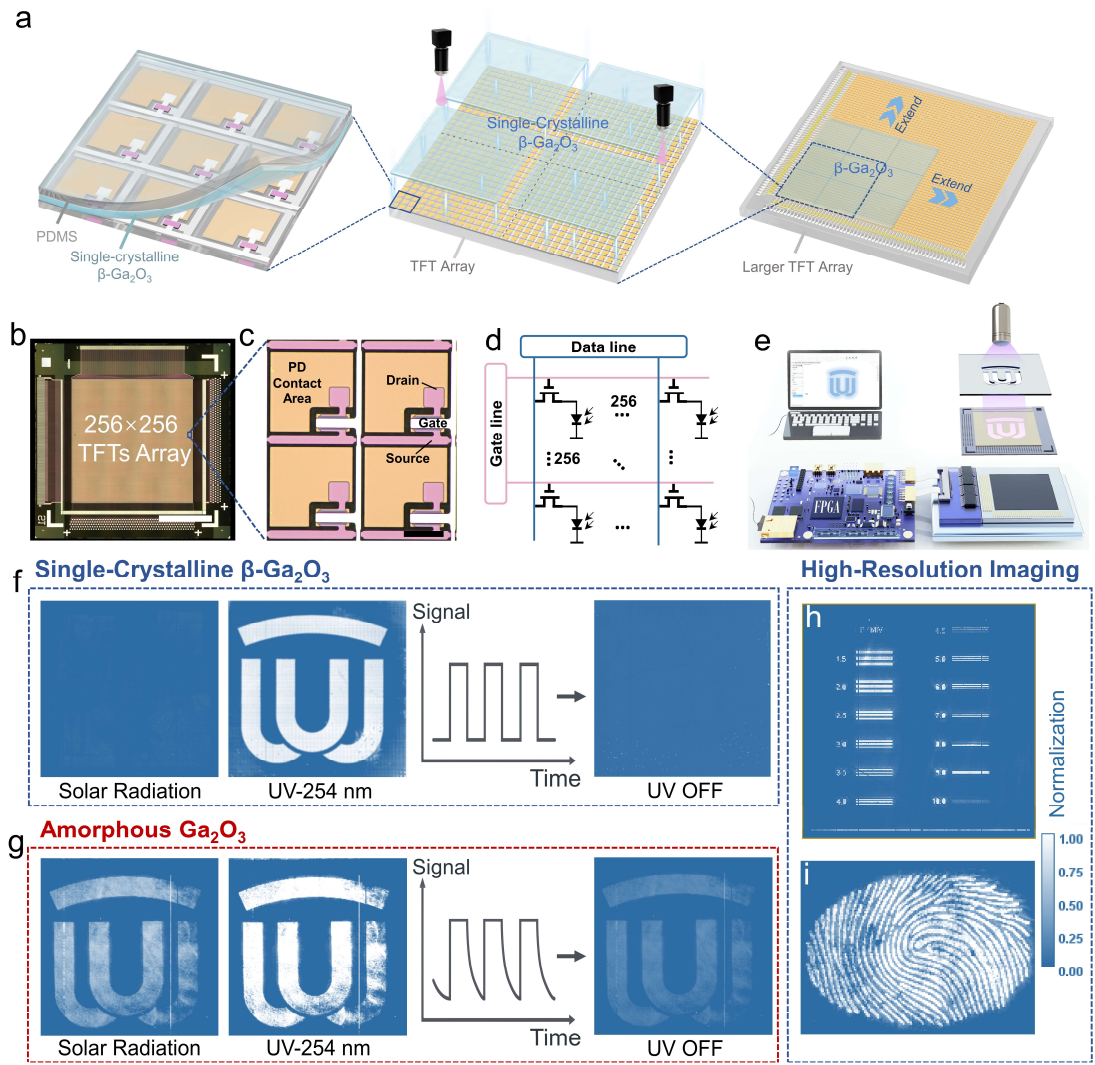

**Fig. 4. Scalable UV imaging arrays enabled by stitching-based membrane assembly. a,** Schematic of fabricating photodetector arrays on high-Resolution TFT backplanes via stitching transfer. **b,** Optical micrograph of the 256×256 α-Si TFT backplane substrate. Scale bar, 1cm. **c,** Optical micrograph of the magnified 2×2 pixel region. Scale bar, 30 μm. **d,** Equivalent circuit diagram of the high-resolution photodetector array. Real-time imaging comparison between single-crystalline β-Ga$_2$O$_3$ **(f)** and amorphous Ga$_2$O$_3$ **(g)** photodetector arrays. Images recorded under solar radiation, under 254-nm UV illumination, schematic illustration of the image-lag mechanism, and residual images acquired several seconds after switching off the UV light. High-resolution imaging of single-crystalline β-Ga$_2$O$_3$ photodetectors: resolution line-pair chart imaging **(h)** and fingerprint imaging **(i)**.

## Conclusion

In summary, we have demonstrated a membrane-enabled electronics platform based on transferable single-crystalline β-Ga$_2$O$_3$ that bridges intrinsic material advantages with scalable electronic integration. By exploiting freestanding single-crystalline membranes and a vertical device geometry, we achieve UV photodiodes with fast, non-persistent photoresponse, high solar-blind selectivity and robust operation, establishing a clear device performance baseline free from parasitic transport effects. Beyond individual devices, we introduce a stitching-based membrane assembly strategy that decouples array resolution from growth wafer size, enabling the scalable construction of high-resolution photodetector arrays on silicon thin-film-transistor backplanes. At the system level, the non-persistent response of single-crystalline membranes directly translates into improved imaging fidelity, eliminating image lag and residual artefacts that commonly limit amorphous wide-bandgap photodetector arrays.

More broadly, this modular integration paradigm offers a pathway to translate the intrinsic advantages of high-quality single-crystalline semiconductors into scalable electronic systems without relying on monolithic wafer growth. By shifting system scalability from the materials growth stage to the backplane and assembly stages, membrane-based integration relaxes long-standing constraints on array size, layout flexibility and heterogeneous integration. Such an approach is particularly relevant for electronic imaging, sensing, and display technologies that demand both high performance and large-area scalability, and may inform future system architectures in which diverse semiconductor membranes are assembled onto common electronic platforms to realize multifunctional and reconfigurable electronic systems.

## Methods

### Epitaxial Growth of Single-crystalline β-Ga$_2$O$_3$ membranes

The Epitaxial β-Ga$_2$O$_3$ membranes were deposited via plasma-assisted molecular beam epitaxy (PA-MBE) under gallium-rich conditions, with the substrate temperature maintained at 650 °C throughout the growth process. In situ reflection high-energy electron diffraction (RHEED) was employed to continuously assess the crystalline quality and monitor film thickness in real time. The substrates used were unintentionally doped (UID) β-Ga$_2$O$_3$ with (100) orientation, sourced from the China Electronics Technology Group Corporation 46th Research Institute. Before epitaxy, the substrates were subjected to a standard organic cleaning procedure, which involved sequential ultrasonic cleaning in acetone, isopropanol, and deionized water, each step lasting 5 minutes. Subsequently, the substrates were dried with a stream of dry nitrogen for 15 minutes to ensure a contamination-free surface.

### PVD Growth of Amorphous β-Ga$_2$O$_3$ Films

Amorphous β-Ga$_2$O$_3$ films were deposited at room temperature by RF magnetron sputtering (ULVAC, CS-200z system). A high-purity Ga$_2$O$_3$ ceramic target was sputtered in an Ar/O$_2$ mixed atmosphere with a total gas flow of 100 sccm. The chamber was evacuated by a combined molecular and mechanical pump to a base pressure of ~5×10$^{-5}$ Pa. During deposition, an RF power of 300 W was applied to the Ga$_2$O$_3$ target. UV-transparent c-plane sapphire wafers were used as substrates after standard solvent cleaning. After sputtering, the samples were naturally cooled to room temperature under vacuum.

**Materials characterizations.**

Optical microscopy (OM) images of samples and devices were obtained using an optical microscope (DSX1000, Olympus).

Atomic force microscopy (AFM) (Dimension ICON, Bruker) was used to characterize the surface morphology of the films, employing non-metal-coated probes (RTESP-300, Bruker) in standard tapping mode.

Scanning Electron Microscopy (SEM) (Gemini500, Zeiss) was used to characterize the surface morphology of the β-$Ga_2O_3$ films at an accelerating voltage of 5 kV.

X-ray Diffraction (XRD) (D8, Bruker) was used to measure 2θ-ω scan and rocking curve with a Cu target.

X-ray photoelectron spectroscopy (XPS) (Nexsa G2, Thermo Scientific) was performed on a Thermo Fisher Scientific ESCALAB Xi+ system equipped with a monochromatic Al Kα X-ray source (1486.6 eV) and a delay-line detector. All measurements were carried out under ultra-high vacuum conditions (~$10^{-9}$ mbar). High-resolution spectra were recorded at a pass energy of 30 eV. Charge compensation was applied throughout data acquisition, and the binding energy scale was calibrated by setting the C 1s peak to 284.8 eV.

UV–vis absorption spectra were obtained using a Shimadzu 3600 UV–vis spectrophotometer with a wavelength range of 200 to 800 nm.

High-angle annular dark-field scanning transmission electron microscope (HAADF-STEM) images were obtained using a spherical aberration-corrected transmission electron microscope (Thermo Fisher Scientific Spectra Ultra) operating at 300 kV. The probe convergence semi-angles were set to 15 mrad and 21 mrad, while the annular detector collection angle ranged from 24 to 121 mrad and from 69 to 200 mrad, respectively. All low-order aberrations were corrected to acceptable levels, achieving a spatial resolution of approximately 0.75 Å.

High-Resolution Transmission Electron Microscopy (HRTEM) were performed using a Thermo-Fisher Talos F200X G2 instrument operating at 200 kV, and Energy-Dispersive X-ray Spectroscopy (EDS) mappings were conducted by a Thermo-Fisher Super X.

The cross-sectional TEM sample was prepared using focused ion beam (FIB) milling with a Thermo Fisher Scientific Helios 5 UX system. Carbon (C) and platinum (Pt) were sequentially deposited on the sample surface to protect the films during the preparation process.

**Exfoliation of Single-Crystalline β-$Ga_2O_3$ membranes**

After epitaxial growth, metal stressor layers were deposited onto the surface of the β-$Ga_2O_3$ membranes using a magnetron sputtering system. A 10 nm Ti adhesion layer was first deposited, followed by a 1 μm Cu layer to provide the required stress, while ensuring a clean and smooth Cu surface. Subsequently, the β-$Ga_2O_3$ membrane together with the deposited metal stressor layer was simultaneously delaminated using thermal release tape (Nitto).

**Fabrication of Membrane-Enabled Vertical β-$Ga_2O_3$ Photodiodes**

After exfoliation, the layer stack from top to bottom consists of the tape substrate, the Cu/Ti metal layers, and the single-crystalline β-Ga$_2$O$_3$ membrane. Subsequently, electrode arrays were fabricated on the exposed surface of the 200 nm freestanding β-Ga$_2$O$_3$ membrane, thereby naturally yielding a vertical device architecture. Specifically, the photosensitive electrode regions were first defined, and a 10 nm Pt layer was deposited through a physical shadow mask using an ion sputtering deposition system (150TS, EMS). Subsequently, a second shadow mask was aligned under an optical microscope, followed by the deposition of a 50 nm Au layer to form probing pads for electrical measurements.

**Transfer Process of Single-Crystalline β-Ga$_2$O$_3$ Membranes**

After exfoliation, a PDMS stamp was first laminated onto the β-Ga$_2$O$_3$ membrane. The entire sample was then placed on a hot plate at 80 °C for 2 min to release the thermal release tape from the underlying Cu metal layer. Subsequently, wet chemical etching was carried out using FeCl$_3$ and buffered oxide etchant (BOE) solutions to remove the metal layers. The sample was then sequentially rinsed in a mixed solution of ethanol and deionized water for 5 min and dried using a nitrogen gun. At this stage, the sample consisted of a PDMS stamp supporting the β-Ga$_2$O$_3$ membrane. The cleaned sample was brought into intimate contact with the target substrate and heated at 60 °C. The PDMS stamp was then peeled off using a stepper stage at a speed of 4000 pulses, thereby releasing and transferring the β-Ga$_2$O$_3$ membrane onto the substrate. Finally, the transferred sample was subjected to Ar plasma cleaning (200 W, 5 min) to remove organic residues introduced during the transfer process.

**Electrical Characterization of the Photodetectors**

A custom-built 254 nm LED coupled into an optical fiber was used as the light source for photoresponse measurements. The electronic properties of I–V and I–t curves were measured using a semiconductor parameter analyzer (4200A-SCS, Keithley). The wavelength-dependent photoresponse in the 200–400 nm range was conducted by a spectral measurement system (DSR-OS-X150A-ZKDDZ, Zolix). A KrF excimer laser (248 nm, Coherent) with a pulse width of 20 ns at a repetition rate of 1 Hz was applied as the excitation source during the temporal response measurements. The temporal response was recorded using an oscilloscope (MSO5000, RIGOL) after signal amplification by a low-noise current preamplifier (Stanford Research Systems SR570).

**Fabrication of the α-Si TFT Array**

The α-Si transistor array was fabricated using a standard TFT process. The process began with the deposition of a 300 nm molybdenum (Mo) gate electrode layer via sputtering on a pre-cleaned glass substrate, which was subsequently patterned using standard photolithography and wet etching. A silicon nitride (SiN$_x$) gate-insulating layer (370 nm) and a hydrogenated α-Si active layer (180 nm) were then sequentially deposited on the bottom gate electrode by plasma-enhanced chemical vapor deposition (PECVD). The channel region was defined using standard photolithography followed by dry etching. Next, the source and drain electrodes, also composed of 300 nm Mo, were deposited and patterned via photolithography and wet etching, resulting in a channel region with a width-to-length ratio (W/L) of 26 μm / 4 μm. To protect the transistor from ambient environmental influences, a 200 nm $SiN_x$ passivation layer was deposited by PECVD. Via holes were opened in this passivation layer using photolithography and dry etching processes. And a 55 nm Mo film was sputter-deposited and patterned to form the pixel bottom electrode.

**Fabrication of the Photodetector Imaging Array**

Using the aforementioned transfer method, β-$Ga_2O_3$ membrane segments, each with a lateral size of 1 cm × 1 cm, were sequentially transferred and assembled under a high-magnification optical microscope using a precision XYZθ alignment stage to align with the pixel regions of the target TFT backplane (see Supplementary Fig. 11). Through lateral stitching of the individual membrane segments, a continuous photodetector array with an effective contact area of 2 cm × 2 cm was achieved. After alignment, a 30 nm ultrathin common top Pt electrode was deposited over the pixel area through a shadow mask using an ion sputtering deposition system (150TS, EMS).

## Author Contribution

Material growth was carried out by T.J. and H.W. Physical property characterization was conducted by X.X., Y.M., and Y.X. Optical characterization of the materials was performed by X.X. and Z.Z., with transient optical measurements guided by K.C. Electron microscopy was performed by H.Z. The membrane exfoliation process was developed by H.W., X.X. The membrane transfer process was developed by X.X., J.C., and T.J. Device fabrication was performed by X.X. Electrical characterization of the devices was conducted by H.H. and X.X., with assistance from Q.H., Y.C., and Z.F.Z. Imaging measurements of the photodetector arrays were performed by Q.H. and X.X. The manuscript was co-written by X.X. and W.K. The research was supervised by W.K., X.Z., B.Z., T.J., and S.L. All authors contributed to discussions and commented on the manuscript. X.X., H.H., Q.H., and H.W. contributed equally to this work.

## Acknowledgments

This work was supported by the National Natural Science Foundation of China (Grant Nos. 62574168 and 62574169), the National Key Research and Development Program of China (2024YFA1208800 and 2023YFB3610200) funded by MOST, the National Natural Science Foundation of Anhui Province (Grant No. 2508085MF145), and the Key Project of Westlake Institute for Optoelectronics (No. 2023GD004). We thank the Westlake Center for Micro/Nano Fabrication, the Instrumentation and Service Center for Physical Sciences (ISCPS), and the Instrumentation and Service Center for Molecular Sciences (ISCMS) at Westlake University for facility support and technical assistance.

## Competing interests

The authors declare no competing interests.

## Additional information

**Correspondence and requests for materials** should be addressed to Tong Jiang, Bowen Zhu, Xiaolong Zhao and Wei Kong.

Supporting Information for

# Scalable Solar-Blind Imaging Enabled by Single-Crystalline β-Ga$_2$O$_3$ Membranes on Silicon Backplanes

Xiang Xu[1, 2, 8], Hong Huang[3, 8], Qi Huang[6, 8], Hao Wang[2, 8], Huaze Zhu[2], Junwei Cao[1, 2], Zheng Zhu[7], Yaqin Ma[1, 2], Yiyang Xu[1, 2], Zhongfang Zhang[4], Yitong Chen[4], Ke Chen[7], Tong Jiang[2, 4]*, Bowen Zhu[4, 5, 6]*, Xiaolong Zhao[3]*, Shibing Long[3], Wei Kong[2, 4, 5, 6]*

[1] School of Materials Science and Engineering, Zhejiang University, Hangzhou, Zhejiang 310027, China.

[2] Department of Material Science and Engineering, School of Engineering, Westlake University, Hangzhou 310000, China.

[3] School of Microelectronics, University of Science and Technology of China, Hefei, Anhui 230026, China.

[4] Zhejiang Key Laboratory of 3D Micro/Nano Fabrication and Characterization, School of Engineering, Westlake University, Hangzhou, Zhejiang 310030, China.

[5] Research Center for Industries of the Future, Westlake University, Hangzhou, Zhejiang 310030, Zhejiang 310030, China.

[6] Westlake Institute for Optoelectronics, Fuyang, Hangzhou, Zhejiang 311400, China.

[7] Center for Neutron Science and Technology, Guangdong Provincial Key Laboratory of Magnetoelectric Physics and Devices, State Key Laboratory of Optoelectronic Materials and Technologies, School of Physics, Sun Yat-Sen University, Guangzhou, Guangdong 510275, China.

[8] These authors contributed equally.

*Email:

jiangtong@westlake.edu.cn; zhubowen@westlake.edu.cn; xlzhao77@ustc.edu.cn; kongwei@westlake.edu.cn

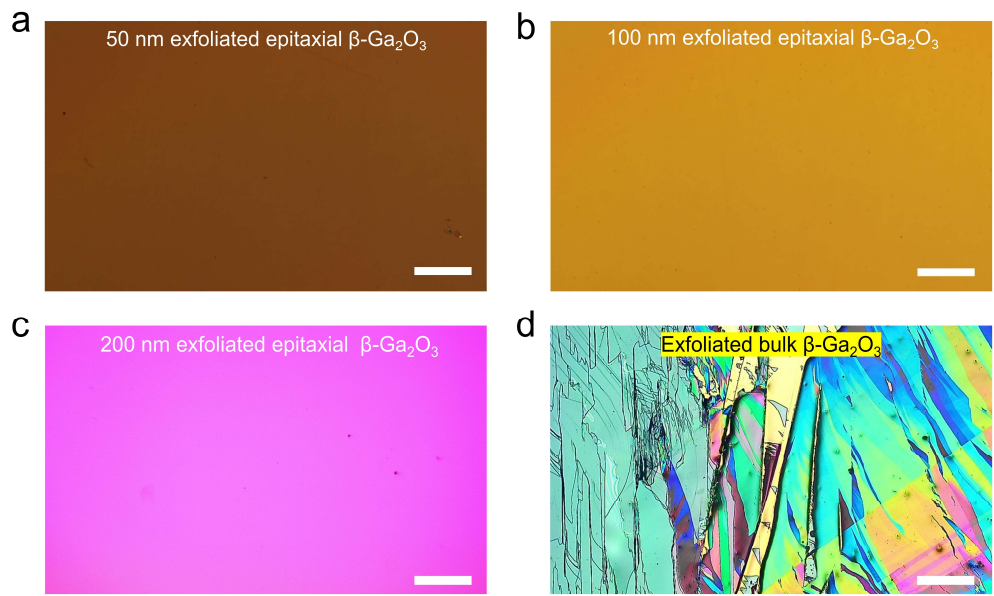

**Supplementary Fig. 1 | Thickness-controlled exfoliation of single-crystalline β-Ga₂O₃ membranes. a-c,** Optical microscopy images of exfoliated single-crystalline epitaxial β-Ga₂O₃ membranes with various thicknesses. Scale bar, 500 μm. **d,** Optical microscopy images of exfoliated bulk β-Ga₂O₃ substrate. The film is directly exfoliated from bulk β-Ga₂O₃ with random thickness. Scale bar, 500 μm.

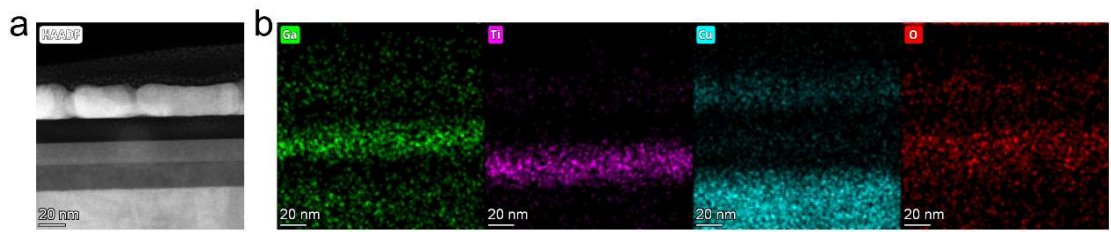

**Supplementary Fig. 2** | Cross-sectional HAADF-STEM image **(a)** and EDS mapping **(b)** of the exfoliated single-crystalline epitaxial β-Ga$_2$O$_3$ membranes on the mental substrate. Scale bar, 20 nm.

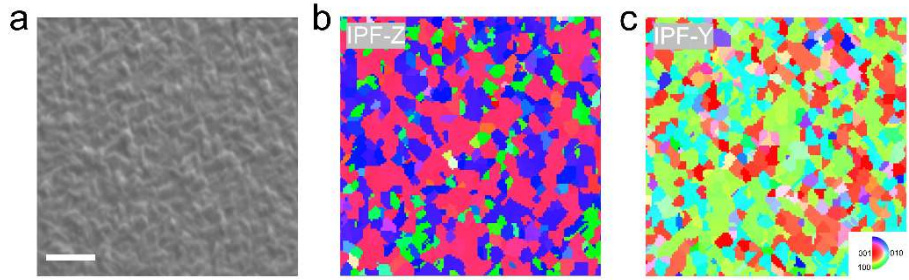

**Supplementary Fig. 3** | Figure **(a)** displays the SEM image acquired during EBSD measurement, while Figure **(b-c)** presents the corresponding EBSD orientation maps of the β-Ga$_2$O$_3$ membrane along the Z (sample normal) and Y (in-plane) directions. The thin film, heteroepitaxially grown on a *c*-plane sapphire substrate via MBE, exhibits a dominant pink contrast across the majority of the mapped area in Figure **(b)**. This color uniformity confirms that the sample possesses a high degree of crystallinity integrity with a strong $(\bar{2}01)$ out-of-plane preferred orientation, consistent with the epitaxial relationship of the substrate. The appearance of other colors indicates that the film is not single-crystalline but exhibits polycrystalline characteristics. Scale bar, 1 μm.

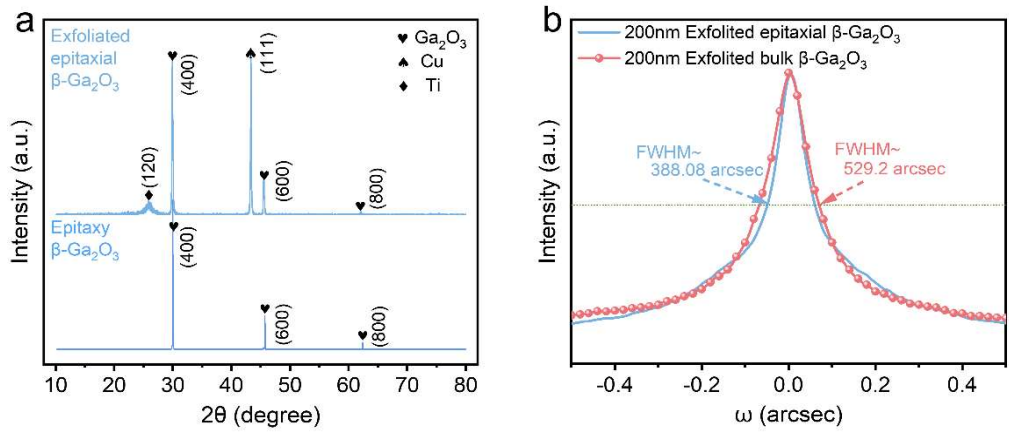

**Supplementary Fig. 4 | a,** 2θ–ω diffraction scans of the epitaxial membrane grown on a β-Ga$_2$O$_3$(100) substrate and after being exfoliated from the epitaxial substrate. **b,** Comparative rocking curves of the (400) reflection: the exfoliated β-Ga$_2$O$_3$ epitaxial membrane versus the film directly exfoliated from the β-Ga$_2$O$_3$ (100) substrate.

**Supplementary Note 1 |** X-ray diffraction (XRD) 2θ–ω scans show identical (400), (600) and (800) reflections for the as-grown epitaxial layer and the freestanding membrane (Supplementary Fig. 4a), indicating the absence of lattice distortion or phase transformation before or after release. The rocking-curve full width at half maximum (FWHM) of the exfoliated epitaxial membrane (388.08 arcsec) is substantially narrower than that of a membrane exfoliated directly from a bulk β-Ga$_2$O$_3$ substrate (529.2 arcsec) (Supplementary Fig. 4b), reflecting reduced mosaic spread and improved crystalline coherence enabled by high-quality homoepitaxial growth.

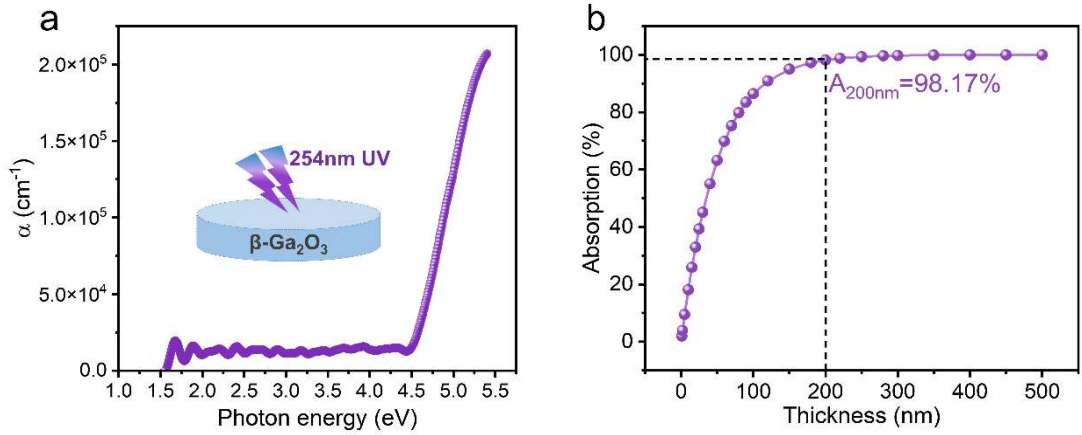

**Supplementary Fig. 5 | a,** Absorption coefficient of single-crystalline β-Ga$_2$O$_3$ epitaxial membrane as a function of photon energy. **b,** The material exhibits strong absorption (>98% for a 200-nm membrane) at the 254 nm UV wavelength.

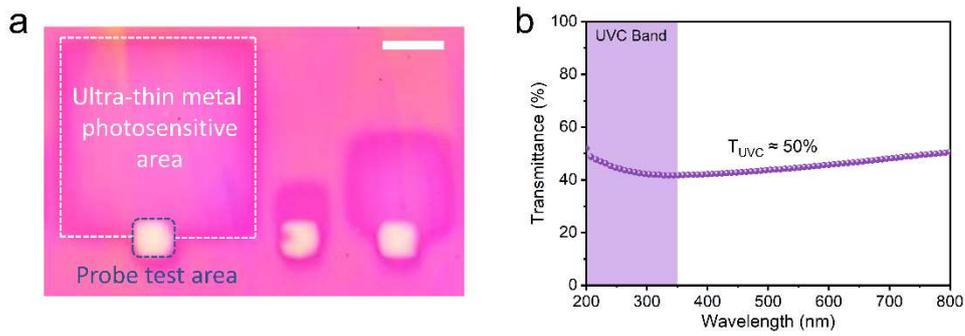

**Supplementary Fig. 6 | Ultrathin metal electrodes in photodetectors. a,** Optical microscopy image showing the surface of the single-crystalline β-Ga$_2$O$_3$ vertical photodetectors. Scale bar, 100 μm. **b,** UV-range transmittance of the ultrathin transparent Pt electrode with a thickness of 10 nm.

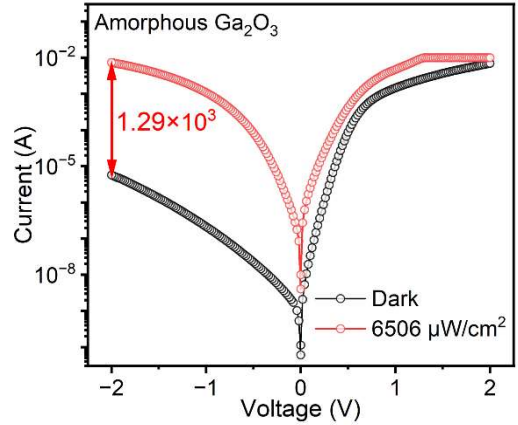

**Supplementary Fig. 7** | Semi-log scale I-V curves of the vertical a-Ga$_2$O$_3$ photodiode measured in the dark and under 254 nm UV illumination (P = 6506 μW/cm$^2$).

**Supplementary Note 2 |**

As shown in Supplementary Fig. 7, the amorphous Ga$_2$O$_3$ photodiode exhibits a clear rectifying behavior in the dark condition. However, it also shows a considerable forward leakage current, reaching 5.72 μA at -2 V. It indicates that the abundant surface states in the amorphous Ga$_2$O$_3$ layer induce strong Fermi-level pinning at the metal–semiconductor interface. The degraded Schottky contact of the device further increases the leakage current. Under DUV illumination, the device exhibits a large photocurrent under reverse bias, achieving a PDCR of 1.29 × 10$^3$ at −2 V. No discernible open-circuit voltage is measured. These observations indicate that the photoresponse does not arise from conventional photocarrier separation and collection but is instead governed by the filling of defect states at the metal–semiconductor interface, which lowers the Schottky barrier under illumination.

**Supplementary Note 3 |**

The responsivity–optical power (R–P) characteristics reveal a minimum detectable optical power of 17 nW cm$^{-1}$, enabled by the low dark current level. In addition, the devices exhibit an almost constant responsivity of ~0.08 A W$^{-1}$ over the measured illumination range, indicating excellent linearity of the photoresponse. Correspondingly, the apparent linear dynamic range (LDR) reaches 118 dB, placing these devices among the highest-performing Ga$_2$O$_3$ photodetectors reported to date.

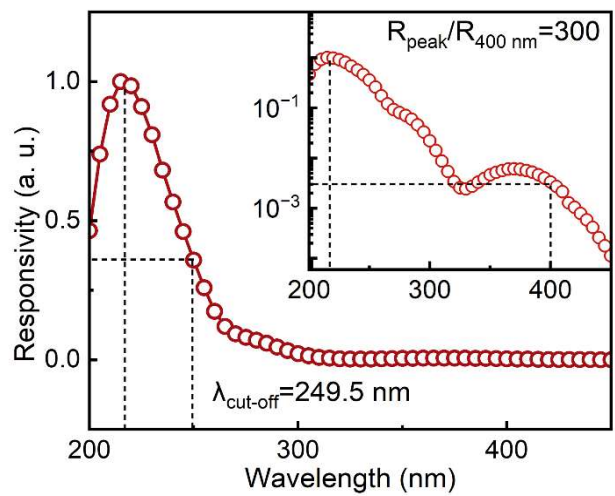

**Supplementary Fig. 8 |** Normalized spectral responsivity of the vertical amorphous Ga$_2$O$_3$ photodiode plotted on a linear scale. The corresponding semi-log plot is shown in the inset.

**Supplementary Note 4 |**

As shown in Supplementary Fig. 8, amorphous Ga$_2$O$_3$ photodiode exhibits a cut-off wavelength of 249.5 nm, corresponding to the intrinsic long-wavelength absorption limit of Ga$_2$O$_3$. However, the amorphous film contains a high density of deep-level defects, which introduces pronounced sub-bandgap absorption. As a result, the device shows a UV–visible rejection ratio of only 300. Compared with the vertical single-crystalline Ga$_2$O$_3$ photodiode, compared with the vertical single-crystalline Ga$_2$O$_3$ photodiode, which shares the same device structure, the use of an amorphous photosensitive layer leads to a substantial degradation in spectral selectivity.

| Crystallinity | Device geometry | UV–visible rejection ratio | Response time (s) | Refs. |
|---|---|---|---|---|
| Single-crystalline | Vertical | ~ $10^6$ | $7.5 \times 10^{-7}$ | This work |
| Single-crystalline | Lateral | $2 \times 10^3$ | $2 \times 10^{-2}$ | Ref.[1] |
| | | $9.2 \times 10^4$ | $8.9 \times 10^{-4}$ | Ref.[2] |
| | | $8 \times 10^2$ | $8.6 \times 10^{-4}$ | Ref.[3] |
| | | $3.5 \times 10^3$ | $1.2 \times 10^{-2}$ | Ref.[4] |
| Poly-crystalline | Vertical | $8.1 \times 10^3$ | $1.12 \times 10^{-1}$ | Ref.[5] |
| | | $3 \times 10^4$ | $9 \times 10^{-3}$ | Ref.[6] |
| | | $1 \times 10^3$ | $6.7 \times 10^{-2}$ | Ref.[7] |
| | Lateral | $2.3 \times 10^4$ | $6.28 \times 10^{-1}$ | Ref.[5] |
| | | $4.06 \times 10^3$ | 1.65 | Ref.[8] |
| | | $1.98 \times 10^5$ | $5.5 \times 10^{-1}$ | Ref.[9] |
| | | $3.5 \times 10^4$ | $1.03 \times 10^{-1}$ | Ref.[10] |
| | | $3 \times 10^5$ | $2 \times 10^{-1}$ | Ref.[11] |

| Crystallinity | Device geometry | UV–visible rejection ratio | Response time (s) | Refs. |
|---|---|---|---|---|
| Poly-crystalline | Lateral | $5\times10^5$ | $5\times10^{-3}$ | Ref.[12] |
| | | $5\times10^3$ | $1.45\times10^{-1}$ | Ref.[13] |
| Amorphous | Vertical | $2.2\times10^3$ | $5\times10^{-2}$ | Ref.[14] |
| | | $1\times10^2$ | 3.3 | Ref.[15] |
| | Lateral | $1\times10^5$ | $5\times10^{-1}$ | Ref.[16] |
| | | $1.49\times10^4$ | $8.5\times10^{-2}$ | Ref.[17] |
| | | $5\times10^3$ | $1.8\times10^{-2}$ | Ref.[18] |

**Supplementary Table 1** | Benchmark of pure $Ga_2O_3$-based photodetectors, comparing UV–visible rejection ratio and response time as a function of crystal type (single-crystalline, polycrystalline and amorphous) and device geometry (lateral versus vertical).

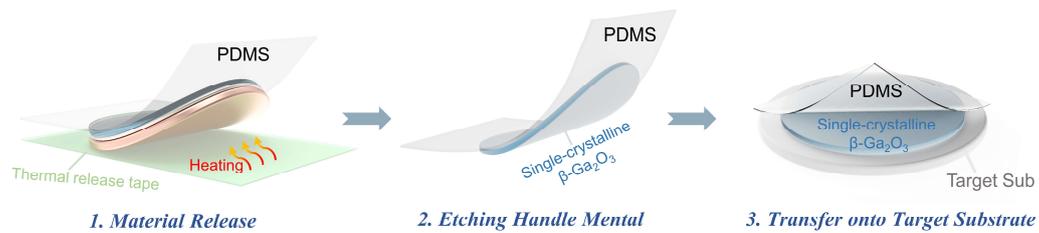

*1. Material Release*    *2. Etching Handle Mental*    *3. Transfer onto Target Substrate*

**Supplementary Fig. 9** | Schematic of the transfer process of the exfoliated single-crystalline β-Ga$_2$O$_3$ membranes. PDMS is first laminated onto the exfoliated material, followed by removal of the thermal release tape and the stress metal layer. The exposed material is then transferred onto the target substrate by releasing the PDMS stamp.

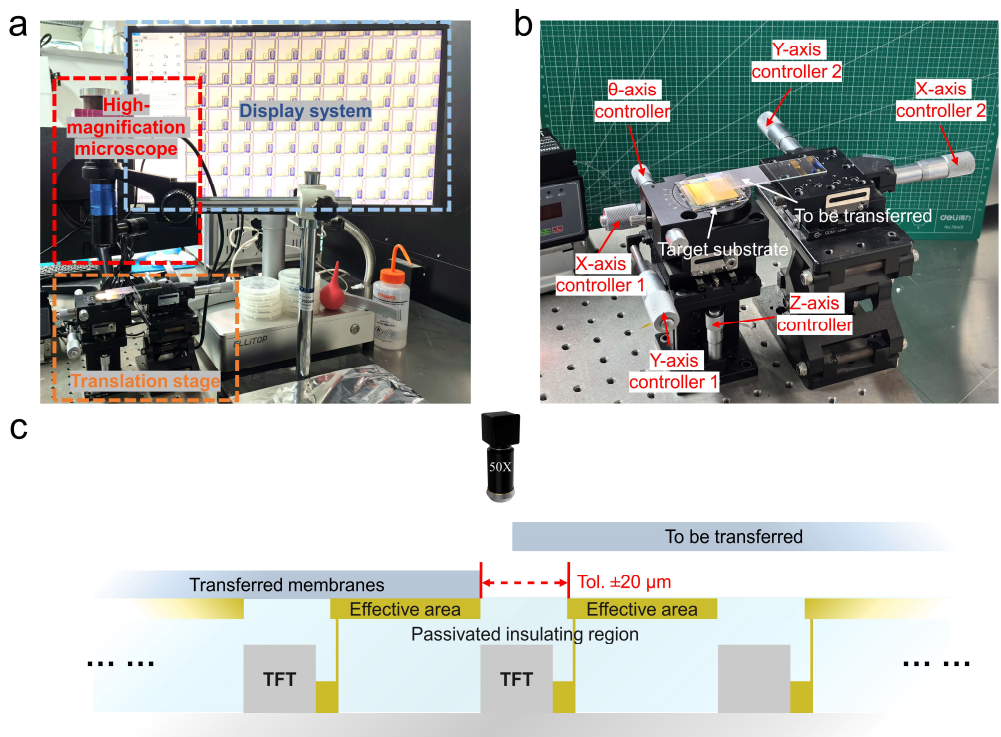

**Supplementary Fig. 10** | Schematic illustration of the alignment and stitching-based transfer system. **a,** Photograph of the material transfer/alignment platform, with the main functional modules highlighted. **b,** Detailed photograph and schematic diagram of the XYZθ four-axis translation/rotation stage. **c,** Schematic of the allowable lateral stitching tolerance (Tol.).

**Supplementary Note 5** |

A stitching-based membrane assembly process was employed to integrate single-crystalline β-$Ga_2O_3$ membranes. Following exfoliation, the freestanding β-$Ga_2O_3$ membrane was first laminated onto a polydimethylsiloxane (PDMS) stamp. The PDMS-supported membrane was then transferred onto a precision XYZθ alignment stage and sequentially aligned to predefined regions on the target substrate under high-magnification optical microscopy. The allowable lateral alignment mismatch between adjacent membrane segments was set within ±20 μm, as the inter-array gap of 20 μm corresponds to a passivated insulating region rather than an electrically active contact area, which is sufficient to ensure seamless membrane stitching without gaps or overlap. By repeating a deterministic pick-and-place procedure, neighbouring membrane segments were laterally stitched together to form a spatially continuous membrane assembly over a large area.

Under typical microscope-assisted transfer conditions, the practical alignment accuracy of the transfer platform is approximately ±1–5 μm, which is well below the stitching tolerance adopted in this work. Owing to the atomically smooth surface and excellent mechanical compliance of the ultrathin membranes, intimate interfacial contact between adjacent membrane segments and the underlying substrate is established via van der Waals interactions, without the use of adhesives or polymer residues. The residual alignment error mainly arises from the limited optical resolution, manual positioning uncertainty, and slight elastic deformation of the PDMS stamp during lamination. It is worth noting that previous studies have shown that by employing a higher precision displacement stage and combining it with optical monitoring, deterministic transfer methods can achieve alignment accuracy at the micrometer or even sub-micrometer level, which points to a clear path for further improving splicing accuracy[19, 20].

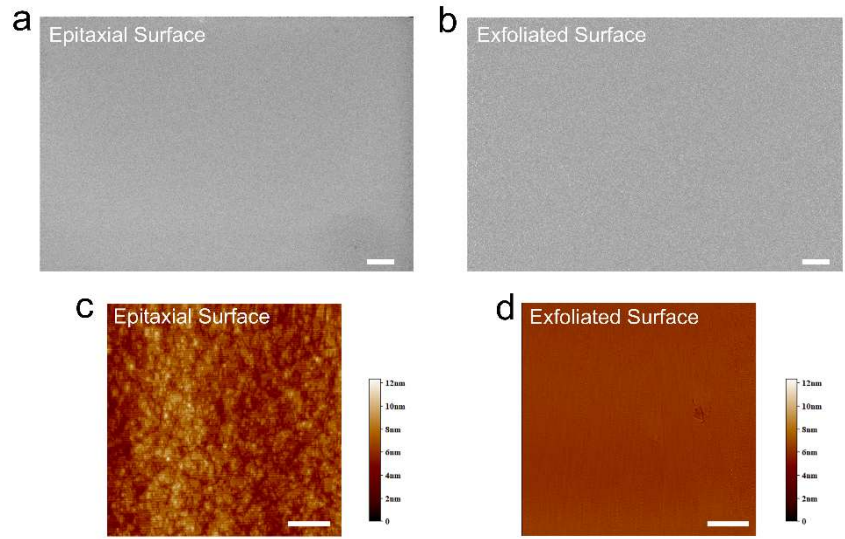

**Supplementary Fig. 11 | Morphological characterization of the interface before and after exfoliation. a,** SEM images of the single-crystalline epitaxial **(a)** and exfoliated **(b)** β-Ga$_2$O$_3$ surface, scale bar, 10 μm. **b,** AFM images of the single-crystalline epitaxial **(c)** and exfoliated **(d)** β-Ga$_2$O$_3$ surface, scale bar, 1 μm.

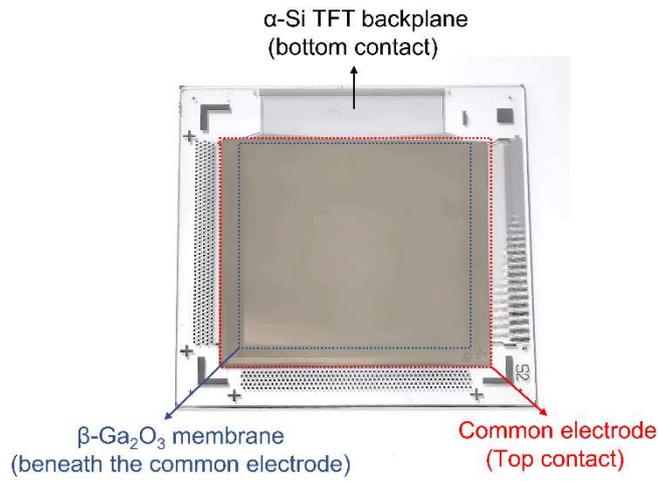

**Supplementary Fig. 12 |** Photograph of the completed vertical single-crystalline β-Ga$_2$O$_3$ photodetector arrays.

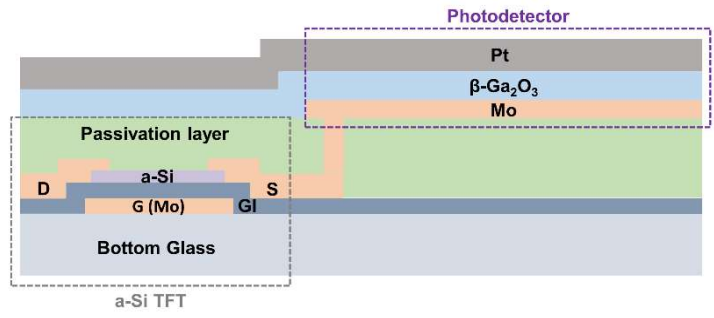

**Supplementary Fig. 13 |** Device schematic of a single photodetector unit.

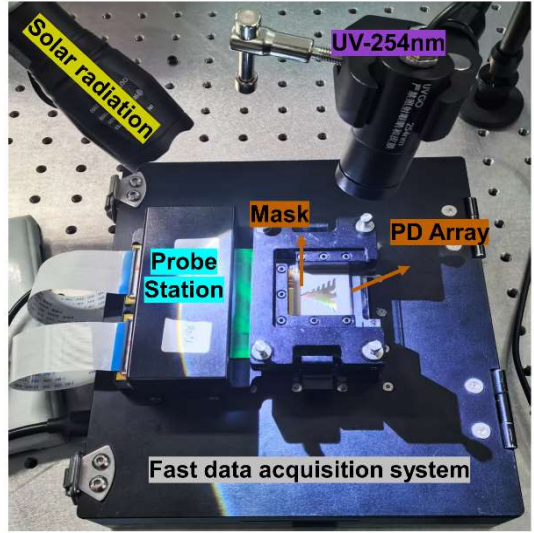

**Supplementary Fig. 14 |** Photograph of the real-time imaging test platform for the photodetector array. This platform allows controlled illumination using simulated solar light and monochromatic 254 nm ultraviolet light while simultaneously recording the output signals from the TFT backplane.

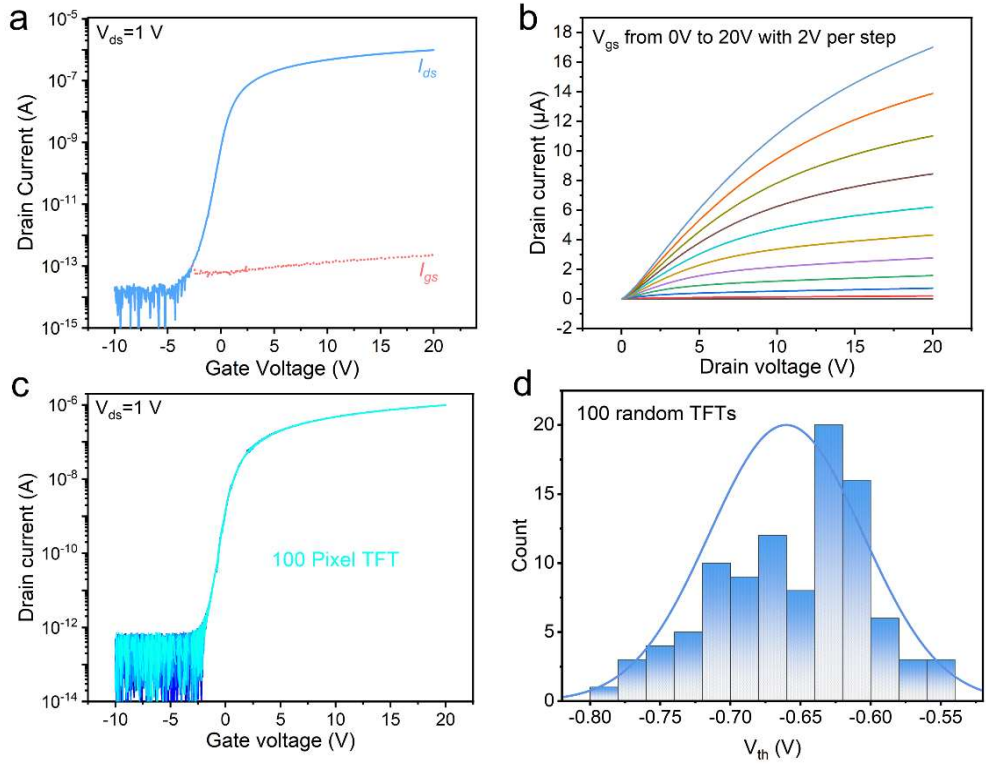

**Supplementary Fig. 15 |** Electrical characterization of the α-Si TFT backplane. **a,** Transfer curve and leakage current of a TFT. **b,** Output characteristics of the TFT. **c,** Transfer curves of 100 randomly selected TFTs. **d,** Statistical distribution of the threshold voltage.

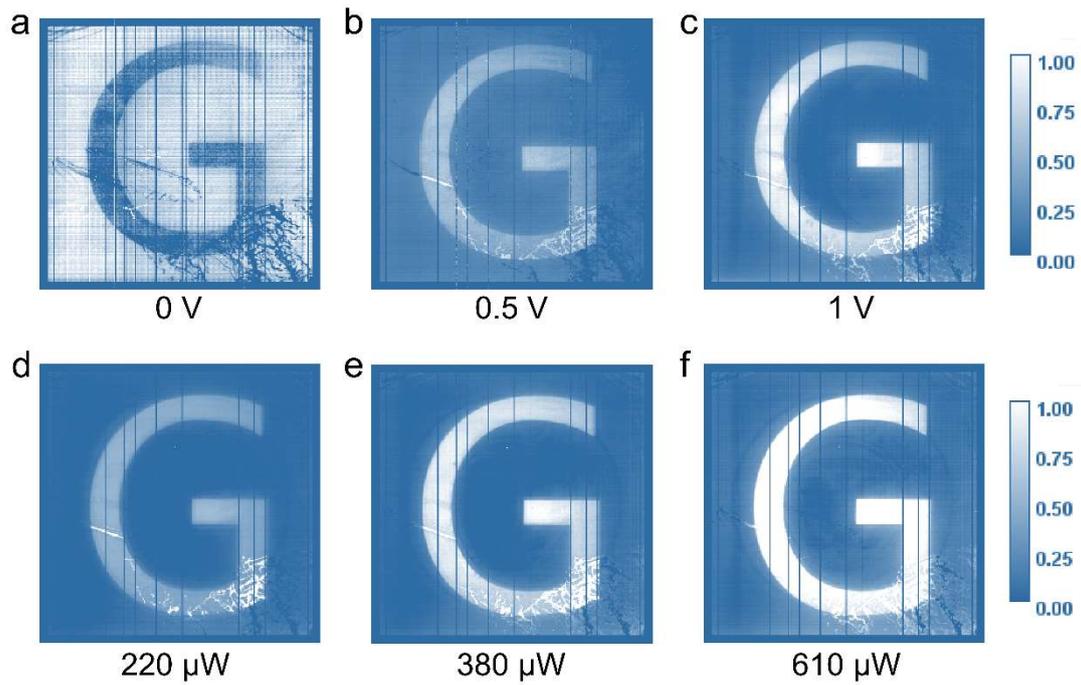

**Supplementary Fig. 16 |** Imaging characteristics of the single-crystalline β-$Ga_2O_3$ photodetector array under different bias voltages and optical power densities. **a–c,** Imaging results recorded at bias voltages of 0, 0.5, and 1 V, respectively. **d–f,** Imaging results obtained under optical power densities of 220, 380, and 610 µW, respectively.